\documentclass[a4paper]{article}

\usepackage{INTERSPEECH2022}
\usepackage{multirow}
\usepackage{caption}
\usepackage{subcaption}
\usepackage{cite}
\usepackage{xcolor}
\title{ An Investigation on Applying Acoustic Feature Conversion \\ to ASR of Adult and Child Speech}
\name{Wei Liu, Jingyu Li, Tan Lee}
\address{Department of Electronic Engineering, The Chinese University of Hong Kong, Hong Kong}
\email{\{louislau\_1129, lijingyu0125\}@link.cuhk.edu.hk, tanlee@ee.cuhk.edu.hk}

\begin{document}

\maketitle
\begin{abstract}
% The performance of child speech recognition is under satisfactory compared to adult speech since only a limited amount of data is available. Employing an automatic speech recognition (ASR) system trained on adult speech to decode the child speech directly will cause severe performance degradation. This is mainly caused by the mismatch between the two domains. In this work, we focus on the adult-to-child conversion on the acoustic feature to alleviate this mismatch. The efficacies of different acoustic feature conversion approaches, either using deep neural network or traditional speech signal processing, are investigated under a fair comparison. The same amount of labeled adult speech is kept to be converted to train the ASR model from scratch. The experiments show that not all of the conversion methods can lead to recognition performance gain. And it is surprising to see the statistic matching, a classic unsupervised domain adaptation approach, takes no effect. It is noted that the F0 distribution of the converted feature set is an important attribute to reflect the conversion quality, while utilizing an adult-child deep classification model to make judgment is shown to be a not appropriate measurement.  

The performance of child speech recognition is generally less satisfactory compared to adult speech due to limited amount of training data. Significant performance degradation is expected when applying an automatic speech recognition (ASR) system trained on adult speech to child speech directly, as a result of domain mismatch. The present study is focused on adult-to-child acoustic feature conversion to alleviate this mismatch. Different acoustic feature conversion approaches, including deep neural network based and signal processing based, are investigated and compared under a fair experimental setting, in which converted acoustic features from the same amount of labeled adult speech are used to train the ASR models from scratch. Experimental results reveal that not all of the conversion methods lead to ASR performance gain. Specifically, as a classic unsupervised domain adaptation method, the statistic matching does not show an effectiveness. A disentanglement-based auto-encoder (DAE) conversion framework is found to be useful and the approach of F0 normalization achieves the best performance. It is noted that the F0 distribution of converted features is an important attribute to reflect the conversion quality, while utilizing an adult-child deep classification model to make judgment is shown to be inappropriate.

%Experimental results reveal that not all of the conversion methods lead to ASR performance gain. Specifically, the approaches of statistic matching and classic unsupervised domain adaptation methods do not show an effectiveness. It is noted that the F0 distribution of converted features is an important attribute to reflect the conversion quality, while utilizing an adult-child deep classification model to make judgment is shown to be inappropriate. 

\end{abstract}

\section{Introduction}
% point the child speech recognition performance is unsatisfied
Automatic speech recognition (ASR) is the technology of converting speech signal into text or equivalent linguistic representation. Deep neural network (DNN) based acoustic model and language model have greatly accelerated the advancement of ASR \cite{hannun2014deep,  zhang2017towards, kim2017joint, wang2019overview, hori2017advances, zeyer2019comparison}, making speech-enabled human-computer interaction feasible and widely accessible. However, the performance of ASR systems for child speech is generally less satisfactory, significantly falling behind state-of-the-art systems for adult speech \cite{claus2013survey,yeung2018difficulties}. This is largely due to the difficulty in collecting sufficient and diverse data of child speech. For ASR research, it is relatively easy to acquire databases of hundreds of hours of adult speech, while child speech databases have much smaller size, e.g., tens of hours, or even are non-existing for many non-major languages.
ASR models trained on a large amount of adult speech data can be used directly to decode child speech utterances. A drastic degradation of recognition performance is expected, because of the mismatch between training and test data in many aspects \cite{shivakumar2020transfer, shahnawazuddin2022robust}.

% Generally, the ASR model trained on large amount of adult speech data can be utilized to decode the child speech directly, while this adult ASR usually suffers a severe performance degradation, because of large mismatch between these two different speech domains \cite{shivakumar2020transfer, shahnawazuddin2022robust}.

% describe what is the mismatch between AC
Child speech was found to present great inter-speaker acoustic variability, which poses a number of modeling challenges \cite{hermansky1990perceptual, lee1999acoustics, shivakumar2022end}. Having a shorter vocal tract than adult\cite{das1998improvements},  children produce speech with higher fundamental frequency (F0) and up-scaled overall formants. The developing articulators of children cause slower and less stable speaking rate \cite{potamianos1997automatic}. Children tend to
commit more pronunciation and grammatical errors during the process of language acquisition \cite{shivakumar2020transfer}. These characteristics all contribute to acoustic mismatch and linguistic mismatch between adult ASR models and child speech. The present study is focused mainly on the acoustic mismatch aspect.

% setting the problem
%Considering a zero-resourced condition for child speech, which means we only have labeled adult speech and unlabeled child speech data, the question is how we can perform better child speech recognition. This is an unsupervised domain adaptation (UDA) setting, domain adversarial training (DAT) is a straightforward UDA method that can be utilized and it operates on the model level \cite{ganin2015unsupervised, sun2018domain}. 
Our goal is to develop an ASR system that can perform optimally on child speech.
We consider a zero-resourced scenario for child speech, i.e., only labeled adult speech and unlabeled child speech data are available. Building an acoustic model by domain adversarial training (DAT) appears to be a straightforward approach in this scenario \cite{ganin2015unsupervised, sun2018domain}.  Here we take another perspective, which is toward reducing the mismatch by transforming input features, e.g., log Mel spectrogram. We aim to convert adult speech features to be like child speech features, and hence carrying acoustic characteristic of child speech. An ASR model trained on the converted features is expected to better match child speech in the acoustic aspect, and hence better recognition performance.

% There are different kinds of choices regarding the conversion approach. In particular, the DNN-based conversion model has attracted much attention in recent years. The most two popular conversion frameworks are auto-encoder disentangle (AE-disentangle) based network and cycle-consistent generative adversarial network (CycleGAN). 
% The AE-disentangle framework considers the speech is coupled with two factors, namely linguistic and para-linguistic \cite{chou2019one}. Usually the linguistic factor refers to the speech content and para-linguistic includes all content irrelevant information, such as speaker identity, emotion, prosody and speaking style. We assume that the information of the difference between adult and child speech is encoded in the para-linguistic part, naming it as a general speaker factor. Ideally, the speech domain conversion can be achieved by modifying this speaker factor.

There are many choices of conversion methods. In particular, the DNN-based conversion models have been prevalent in recent years. In this study, disentanglement-based auto-encoder (DAE) is adopted as a basic framework for investigation \cite{chou2019one, yuan2021improving, li2018disentangled, qian2019autovc}. This framework considers two coupled factors of variation in speech, namely linguistic and para-linguistic factors. The linguistic factor refers to the speech content and the para-linguistic factor covers all content-irrelevant information, including speaker identity, emotion, prosody, and speaking style. We assume that the information about the difference between adult and child speech is encoded as part of the para-linguistic factor and name it as a general speaker factor. Ideally, acoustic feature conversion between adult and child speech can be achieved by modifying this speaker factor.

Apart from the DAE, the cycle-consistent generative adversarial network (CycleGAN) is a popular approach to perform domain transfer. It was proposed for image translation in computer vision \cite{zhu2017unpaired}. Cycle-consistent and adversarial loss are utilized to train CycleGAN without requiring paired data. It has been applied to many unsupervised domain adaptation tasks, including voice conversion \cite{kaneko2018cyclegan, kaneko2019cyclegan, prananta2022effectiveness}. 
In addition to these DNN-based conversion models, traditional signal processing methods, e.g., formant modification \cite{ yeung2021fundamental} and time-scale modification \cite{shahnawazuddin2022robust}, can also be applied to acoustic feature conversion.

% In this paper, the efficacies of different acoustic feature conversion approaches are investigated under a fair comparison, which means we keep the same amount of labeled adult speech to be converted to train the ASR model. Particularly, the AE-disentangle based framework are explored and experimented with various settings in order to better perform adult-to-child conversion.

In this paper, the disentanglement-based AE framework is investigated with comparison to other acoustic feature conversion approaches under a fair setting, i.e., the ASR model is trained with the converted features from the same amount of labeled adult speech. The efficacies of component modules in the DAE framework are investigated through an ablation study.
% The F0 distribution is found to be an important attribute to evaluate the child characteristic of the converted acoustic feature, while the DNN-based adult-child classifier is not an appropriate measurement. 

%The rest of the paper is organized as follows. 
The DAE based conversion framework will be illustrated in Section 2. Section 3 describes the experimental setup of the acoustic feature conversion and ASR model training. Section 4 gives the results and the work is concluded in Section 5.

% referenced processing method
% \begin{itemize}
    %\item CYCLE-GANS FOR DOMAIN ADAPTATION OF ACOUSTIC FEATURES FOR SPEAKER RECOGNITION. We explored the usage of cycle-GANs for learning feature mapping functions across telephone microphone domains without any parallel data between both domains.  The main challenge that we observed was to transfer features to another domain while preserving the structure (like lingusitic information, speaker information, etc). The best results were obtained when the generators were trained to learn a residual mapping between its input and output (cycle-GANs system C).
    %\item Unsupervised Domain Adaptation for Acoustic Scene Classification Using Band-Wise Statistics Matching.
     %First, just before the training phase, the sample mean and standard deviation for each frequency band across every sample in the source-domain training dataset is computed. Second, at inference time, a band-wise standardization is applied to the target-domain test data so to obtain zero-mean and unit-variance frequency bands throughout the dataset. Third, the standardized dataset is finally adapted using the means and standard deviations computed at Step 1.
     %\item Unsupervised domain adaptation for robust speech recognition via variational autoencoder-based data augmentation. Similar to what we operate on the speaker embedding space. Think the average of latent speaker variables can represent its domain, child or adult in this case.
%\end{itemize}

\begin{figure}[t]
  \centering
  \includegraphics[width=\linewidth]{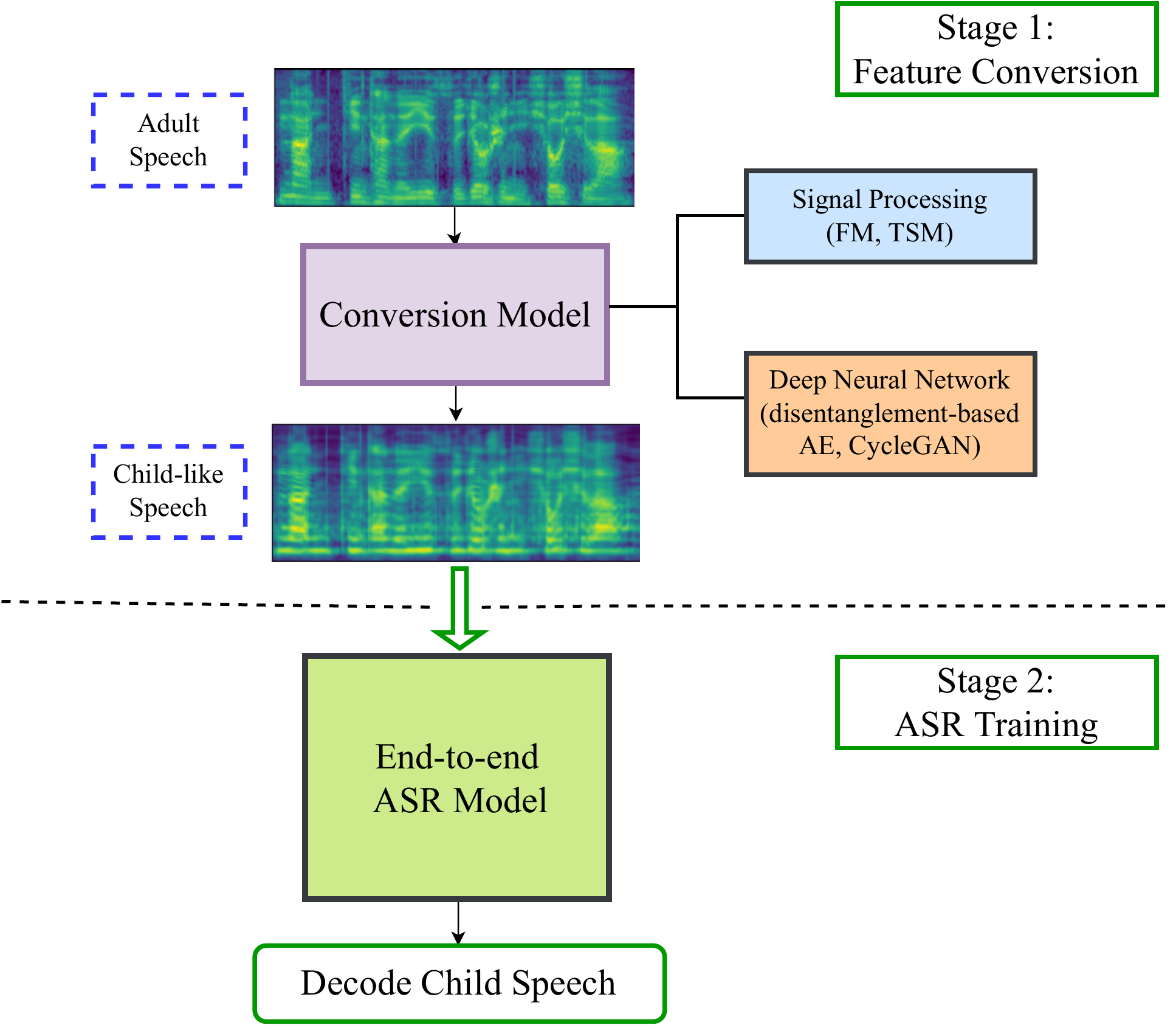}
  \caption{The overall workflow of the adult-to-child acoustic feature conversion for ASR.}
  \label{fig:workflow}
\end{figure}

\section{Methods}
% First, the overall conversion workflow is briefly described. Next, the AE-disentangle conversion framework and its core module AdaIN will be introduced in detail as follows.

\subsection{Workflow of experimental process}
The overall workflow of acoustic feature conversion process includes two steps as illustrated in Figure \ref{fig:workflow}. The first step is to perform adult-to-child feature transformation via a conversion model. The conversion approaches are categorized into two streams: signal processing based methods, e.g., formant modification (FM) and time-scale modification (TSM); DNN based methods, e.g., disentanglement-based AE and CycleGAN. After feature conversion, the second step is to train an ASR model  using the converted speech feature set. The degree of improvement on recognition performance on child test speech is evaluated against systems without using converted features. 

\subsection{AdaIN voice conversion}
Voice conversion (VC) aims to modify content irrelevant information while preserving the linguistic content.
The same idea is adopted  for acoustic feature conversion except that the vocoder for converting spectrogram to waveform is not needed. AdaIN in \cite{chou2019one} is a disentanglement-based auto-encoder network, which performs one-shot VC by separating content and speaker embeddings with instance normalization (IN). Three modules constitute the AdaIN. They are the content encoder $E_{c}$, the speaker encoder $E_{s}$ and the decoder $D$, respectively. The acoustic feature $\mathbf{x}$ is fed into $E_{c}$ and  $E_{s}$ as input. $E_{c}$ generates a sequence of content embeddings $\mathbf{zc}$, while $E_{s}$ produces the speaker representation $\mathbf{zs}$. The decoder outputs $\mathbf{\hat{x}}$ which is intended to reconstruct $\mathbf{x}$ from $\mathbf{zc}$ and $\mathbf{zs}$. The IN is applied to the content encoder to remove speaker-related information, while the adaptive IN \cite{huang2017arbitrary} is used to provide the global speaker information encoded by $E_{s}$ to the decoder. The whole network is trained to minimize the reconstruction loss $|| D(\mathbf{zc}, \mathbf{zs}) - \mathbf{x}||_1$ in an unsupervised manner.
The conversion is performed by feeding $\mathbf{zc}_{src}$ of the source speaker and $\mathbf{zs}_{tar}$ of the target speaker to the decoder, i.e., $D(\mathbf{zc}_{src}, \mathbf{zs}_{tar})$.

\begin{figure}[t]
  \centering
  \includegraphics[width=0.99\linewidth]{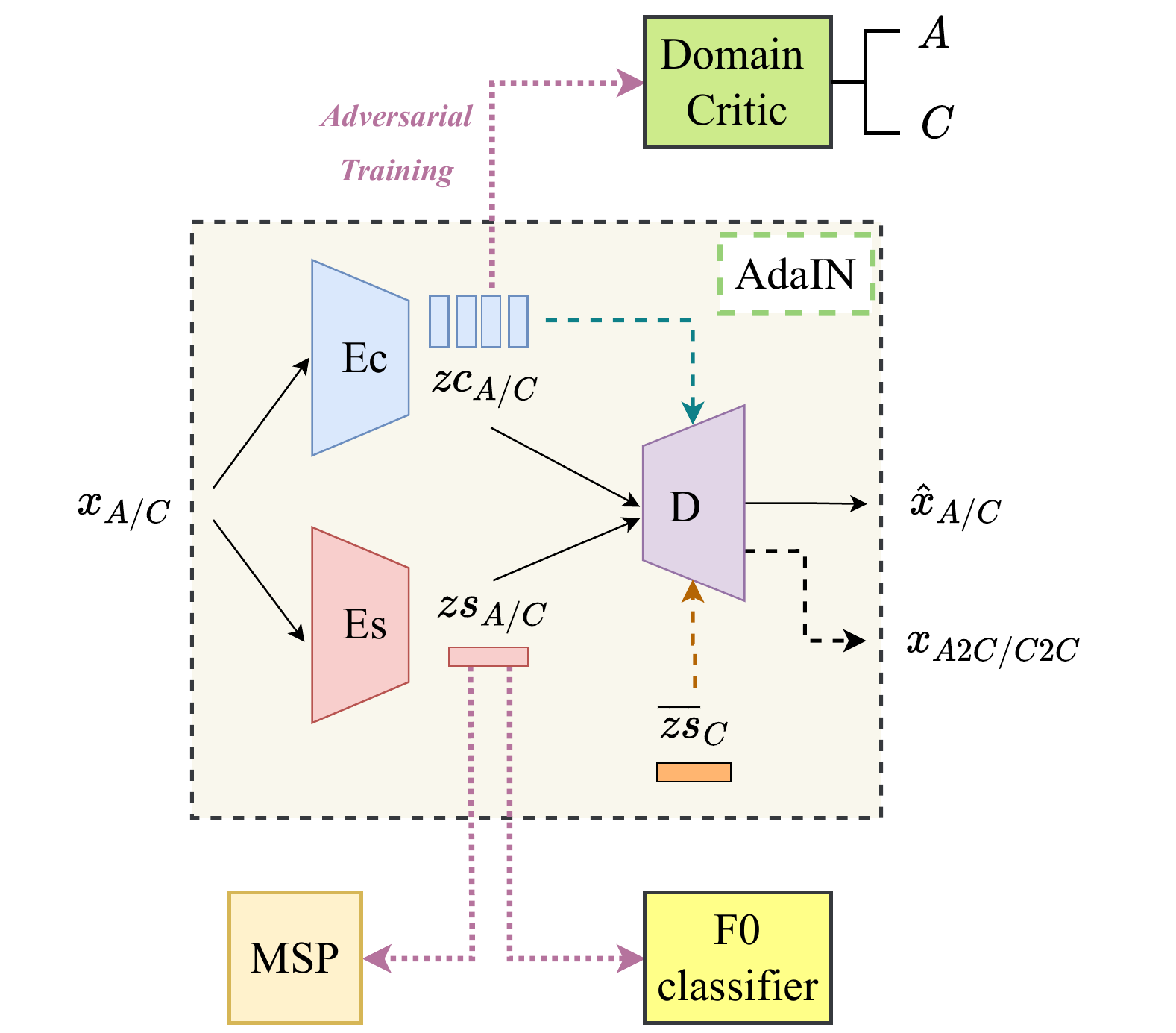}
  \caption{The data flow of the DAE conversion framework to conduct adult-to-child conversion.}
  \label{fig:ae-disentangle}
\end{figure}

\subsection{DAE acoustic feature conversion}
 As shown in Figure \ref{fig:ae-disentangle}, the AdaIN network (in dashed block) is the core module of the DAE framework, which aims to perform adult-to-child conversion in the acoustic feature space.
 The subscript $A/C$ in a variable indicates that it represents the adult or the child domain. The solid-line arrows illustrate the reconstruction process in the training phase, while the dashed arrows refer to the conversion stage. $\mathbf{\overline{zs}}_C$ is the representation of child domain and calculated by averaging all speaker embeddings of child speech utterances. $\mathbf{x}_{A2C}$ denotes the converted acoustic feature, which is generated by replacing $\mathbf{zs}_A$ with $\mathbf{\overline{zs}}_C$. 
 In the evaluation of ASR models trained by $\mathbf{x}_{A2C}$, to overcome the great mismatch between the real and generated speech features, the child test speech is also performed by $C2C$ conversion, and the corresponding generated feature is denoted as $\mathbf{x}_{C2C}$.
 %Since there is great mismatch between the real and generated speech features, we also perform $C2C$ conversion for $\mathbf{x}_C$ and the conversion output $\mathbf{x}_{C2C}$ is regarded as the corresponding child test speech for ASR trained by $\mathbf{x}_{A2C}$. 
 The whole conversion process can be expressed as: 

\begin{equation}
    \mathbf{x}_{A2C/C2C} = D(\mathbf{zc}_{A/C}, \mathbf{\overline{zs}}_C)
\end{equation}

The original AdaIN network is trained in a fully unsupervised manner, i.e., only speech data are required. Nevertheless, the domain class label and median F0 value are used to facilitate a better conversion.
A domain-critic module is built on top of $\mathbf{zc}$, which is adversarially  trained to force content embeddings  to be domain-invariant. Since the F0 distribution is found to be an important attribute to discriminate adult and child speech, the median F0 of each utterance is estimated and used to train the F0 classifier. Moreover, the matrix subspace projection (MSP \cite{li2020latent}) is applied to the speaker embedding space. It transforms $\mathbf{zs}$ into a low-dimensional attribute space, adult versus child domain in this case, which is expected to make $\mathbf{zs}$ contain more discriminative domain information.   

\section{Experimental Setup}
\subsection{Dataset}
Acoustic feature conversion is performed to reduce mismatch between the two speech domains, namely adult speech and child speech. The adult speech ($A$) data are from a subset of the AISHELL1 corpus. The child speech ($C$) data are from the 2021 SLT CSRC (short for Children Speech Recognition Challenge) dataset. AISHELL1 \cite{bu2017aishell} is an open-source dataset of Mandarin speech by adult speakers in reading style. It is intended and widely used for ASR research. The CSRC dataset \cite{yu2021slt} consists of two parts of child speech with different speaking styles. The first part, denoted by $C_1$, contains read speech. The second part, denoted by $C_2$, contains conversational speech. 
The test set of both $C_1$ and $C_2$ contain 2 hours of speech.
% $2.05$ and $2.16$ hours.
The train data sets of AISHELL1 and CSRC are summarized as in Table \ref{tab:dataset}.

\begin{table}[t]
\centering
% \caption{The statistic information of the train sets.}
\caption{A summary of training datasets used in this research.}
\begin{tabular}{c||ccc}
\toprule
Data set   & $A$       & $C_1$      & $C_2$      \\ \hline
Duration (hrs)     & 60      & 24      & 25      \\ 
\# of Utts & 48, 515     & 23, 824 & 25, 447 \\ 
\# of Speakers & 137     & 742     & 133     \\ 
Speaker age & 18-60      & 7-11    & 4-11    \\
Speaking style & read    & read    & conversational \\ \bottomrule
\end{tabular}
\label{tab:dataset}
\end{table}

\subsection{Feature conversion}
% To conduct acoustic feature conversion, 
80-dimensional log Mel spectrograms are extracted from raw audio with 25 ms window length and 10 ms hop length. All audio data are sampled at 16 kHz. The feature sets of $A$, $C_1$ and $C_2$ are pooled together where global mean and variance normalization (GMVN) is applied. The disentanglement-based AE network is trained with the normalized features.
It is noted that the speech in $C_2$ exhibits significant mismatch with $C_1$ in the preliminary ASR experiment. This is related to the speaking styles difference, i.e., read speech vs conversational speech. Therefore, we consider adult-to-child conversion under the same speaking style. Specifically, $C_1$ is regarded as the target child domain of interest, meaning that there are totally three conversion types, namely $A2C_1$, $C_12C_1$, and $C_22C_1$. 

The domain-critic module is implemented with a three-layer fully-connected (FC) network for domain classification on three classes, namely $A$, $C_1$, and $C_2$. The F0 classifier adopts a similar network structure to perform a 10-class task. The utterance-wide median F0 values are divided in 10 equal intervals covering the range from 100 Hz to 350 Hz. F0 estimation is implemented by the Parselmouth library.  When applying MSP on the speaker embedding space, the attribute label is designed as a 2-dimensional vector, in which the first element represents the adult/child domain and the second element represents the read/conversational speaking style. The core AdaIN network follows the same Conv1D layers architecture as described in \cite{chou2019one}. The channel dimension is 512. The variational regularization is not imposed on the content embeddings $\mathbf{zc}$  \cite{kingma2013auto}, though it is applied in the original AdaIN network by default. The model is trained with $segment\_length=128$, $batch\_size=128$ for 100k steps. The optimizer is Adam with $learning\_rate=0.0005$ \cite{kingma2014adam}.

In the conversion stage,  the average of speaker embeddings from the $C_1$ development set (denoted as $\overline{\mathbf{zs}}_{C1}$) is computed to represent the target child domain. The adult-to-child speech feature conversion is performed by replacing the original speaker embedding with $\overline{\mathbf{zs}}_{C1}$.
The decoded output will be de-normalized as the converted spectrogram.

% Simply introduce MaskCycleGAN and F0-based formant modification used as conversion method...
Apart from the disentanglement-based AE framework, other conversion methods are  evaluated in our experiments. The MaskCycleGAN network \cite{kaneko2021maskcyclegan} is utilized to perform direct domain transformation from $A$ to $C_1$. Two generative models and two discriminative models are trained with 20k utterances for each domain. In terms of non-DNN methods, F0-based feature normalization conducts the formant modification by assuming a linear relation between F0 and formants on the Mel scale \cite{yeung2021fundamental}. The target value of normalized F0 is empirically set to be 270 Hz. 
%The additional TSM cannot show improvement, so it will not be reported.
In addition, statistic matching on spectrum space (denoted as \textit{Stats}) is evaluated \cite{mezza2021unsupervised}, in which the set $A$'s feature is first normalized by its mean and standard deviation (std), and then de-normalized by the mean and std of the $C_1$ set. The Correlation Alignment (\textit{Coral} \cite{sun2017correlation}) is also experimented to minimize the domain shift by aligning the second-order statistics of $A$ and $C_1$ distributions. The only difference with the \textit{Stats} method is that \textit{Coral} uses the covariance instead of the std.

\subsection{End-to-end ASR}
The speech recognition model used in our experiments adopts the joint CTC-attention architecture \cite{kim2017joint}. It consists of three components, namely the shared encoder, the attention decoder, and the CTC loss layer. The encoder comprises 12 \textit{Conformer} layers and the decoder has 6 \textit{Transformer} layers \cite{vaswani2017attention}. The input feature of ASR model is 80-dimensional log Mel spectrogram, either the original or the converted one. The ASR model is trained to minimize the weighted summation of the attention decoder loss and CTC loss, where the CTC loss weight is empirically set to be 0.3. The maximum number of epochs is set to be 150 for model convergence on 60-hour training data. The SpecAug \cite{park2019specaugment} and GMVN are applied by default. The language model is disabled if not stated otherwise. The ASR experiments are conducted using the ESPnet toolkit \cite{watanabe2018espnet}.
%The language model is disabled to focus on the effect of the change of acoustic attribute if not specifically state.    

\begin{table}[t]
\centering
\caption{The WER (\%) results on child test speech using ASR trained by different converted acoustic features and unconverted counterparts.}
\begin{tabular}{ccc|cc}
\toprule
\multicolumn{1}{c|}{\multirow{2}{*}{Type}}                                                          & \multicolumn{1}{c|}{\multirow{2}{*}{Model}} & \multirow{2}{*}{\begin{tabular}[c]{@{}c@{}}Train\\ set\end{tabular}} & \multicolumn{2}{c}{Test set} \\ \cline{4-5} 
\multicolumn{1}{c|}{}                                                                               & \multicolumn{1}{c|}{}                       &                                                                      & $C_1$     & $C_2$            \\ \hline  \hline
\multicolumn{1}{c|}{\multirow{2}{*}{\begin{tabular}[c]{@{}c@{}}Without \\ Conversion\end{tabular}}} & \multicolumn{1}{c|}{Baseline}               & \multirow{2}{*}{$A$}                                                 & 29.0      & 76.6             \\
\multicolumn{1}{c|}{}                                                                               & \multicolumn{1}{l|}{+ LM  (C)}              &                                                                      & 25.8      & 71.7             \\ \hline
                                                                                                    &                                             &                                                                      & $C_12C_1$ & $C_22C_1$        \\ \hline
\multicolumn{1}{c|}{\multirow{5}{*}{\begin{tabular}[c]{@{}c@{}}With\\ Conversion\end{tabular}}}     & \multicolumn{1}{c|}{\textit{DAE}}                     & \multirow{5}{*}{$A2C_1$}                                             & 28.5      & 75.1             \\
\multicolumn{1}{c|}{}                                                                               & \multicolumn{1}{c|}{\textit{CycleGAN}}               &                                                                      & 31.7      & \textbackslash{} \\
\multicolumn{1}{c|}{}                                                                               & \multicolumn{1}{c|}{\textit{F0-norm}}                &                                                                      & \textbf{28.3}      & \textbf{74.6}             \\
\multicolumn{1}{c|}{}                                                                               & \multicolumn{1}{c|}{\textit{Stats}}                  &                                                                      & 30.3      & 76.3             \\
\multicolumn{1}{c|}{}                                                                               & \multicolumn{1}{c|}{\textit{Coral}}                  &                                                                      & 30.3      & 77.1            \\ \bottomrule
\end{tabular}
\label{tab:asr_reslut}
\vspace{-2mm}
\end{table}

\section{Results and Analysis}

\subsection{Comparison of different conversion methods}
\begin{figure}[t]
         \centering
         \includegraphics[width=\linewidth]{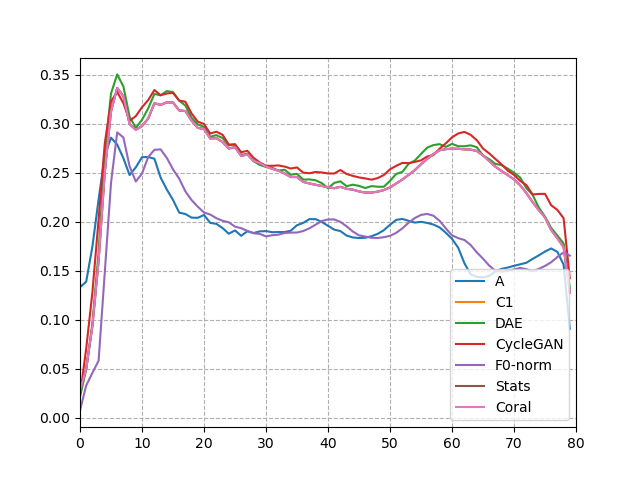}
         \caption{The mean statistic of different acoustic features \\ (converted $A2C_1$ vs original $A$ \& $C_1$).}
         \label{fig:mu_spec}
\vspace{-2mm}
\end{figure}

% (1) DNN-based (2) SP-based (3) Statistic matching
The word error rate (WER) results of child test sets are given in the Table \ref{tab:asr_reslut}. 
%Large mismatch between $C_1$  and $C_2$ can be clearly seen from the baseline, in which the ASR model is trained using the original data of set $A$ and test on $C_1$ and $C_2$ respectively. 
The baseline model trained by original $A$ set attains $8.9\%$ WER on the adult test set. In $C_1$ and $C_2$, the performance of baseline model suffers from great mismatch even with language model (LM).
The ASR models trained with converted acoustic features are compared against the baseline. Five feature conversion methods are evaluated in our experiments.
% The ASR models trained by different types of converted acoustic features are evaluated and compared against the baseline.
The DAE based conversion model (\textit{DAE} for short) achieves $0.5\%$ absolute recognition improvement on the $C_1$ test set, and $1.5\%$ on the $C_2$ test set. The best recognition performance can be attained by the F0 normalization (\textit{F0-norm}). Not all of the conversion methods can  lead to performance gain, e.g., the \textit{CycleGAN}. Although, it is still surprising to observe that the approach of statistic matching  takes no effect and even makes degradation.
%where the Correlation Alignment (\textit{Coral} \cite{sun2017correlation}) leverages the corvariance instead of the std used by \textit{Stats}. 
The mean statistics of the log Mel spectrogram with 80 channels are plotted  in Figure \ref{fig:mu_spec}, in which the \textit{Stats} and \textit{Coral} curves overlap with that of $C_1$. A reasonable formants up-scale is noted in \textit{F0-norm} compared to the original $A$.
%in which the \textit{AE} curve is the closest to that of the $C_1$ and \textit{F0-norm} shows a reasonable formants up-scale compared to the original $A$ set.

To investigate the efficacies of different network components in the \textit{DAE} conversion model, an ablation study is carried out as shown in Table \ref{tab:ablation}. The usage of domain critic and F0 classifier are represented by the symbol DAT and F0\_clf. Having DAT on the content encoder is important for effective disentanglement. 
% which can implicitly force the domain related information flow into the speaker encoder branch.
The role of MSP seems not to be as useful as F0\_clf.
Imposing variational regularization on the content encoder does not work. 
Besides, an experiment with vanilla \textit{DAE} found that additional benefits ($29.1\% \xrightarrow{} 28.8\%$) can be attained by training more steps (200k in this case). 
The limited performance improvements are noted in all cases, which may suggest the quantity of data to train \textit{DAE} is insufficient.

% Please add the following required packages to your document preamble:
% \usepackage{multirow}
\begin{table}[t]
\centering
\caption{The ablation study on the DAE-based framework. The backslash \textbackslash ~represents removing that component(s).}
\begin{tabular}{c|c|cc}
\toprule
\multirow{2}{*}{\begin{tabular}[c]{@{}c@{}}Conversion\\ Model\end{tabular}}             & \multirow{2}{*}{Configuration}                                                                  & \multicolumn{2}{c}{WER (\%) of test set} \\ \cline{3-4} 
                                                                                        &                                                                                          & $C_12C_1$     & $C_22C_1$    \\ \hline
\multirow{8}{*}{\begin{tabular}[c]{@{}c@{}}DAE-based\\framework\end{tabular}}   & All                                                                                      & 28.5          & 75.1         \\ \cline{2-4} 
                                                                                        & \textbackslash ~DAT                                                                       & 28.9          & 75.3         \\
                                                                                        & \textbackslash ~F0\_clf                                                                       & 28.9           & 75.2         \\
                                                                                        
                                                & \textbackslash ~MSP                                                                       & 28.6           & 75.3         \\
                                                                                        & \textbackslash ~MSP \& F0\_clf                                                            & 28.7           & 75.7          \\
                                                                                        & \textbackslash ~DAT \& MSP                                                                & 29.0          & 75.7         \\
                                                                                        & \textbackslash ~DAT \& F0\_clf                                                            & 29.0           & 75.2          \\
                                                                                        & \begin{tabular}[c]{@{}c@{}}\textbackslash ~DAT \& MSP \\ \& F0\_clf (vanilla)\end{tabular} & 29.1          & 75.8         \\ \hline
\multirow{3}{*}{\begin{tabular}[c]{@{}c@{}}vanilla \textit{DAE}\\ + variational\end{tabular}} & KL weight: 1.0                                                                           & 30.2          & 76.7         \\
                                                                                        & KL weight: 0.1                                                                           & 29.2           & 76.4          \\
                                                                                        & KL weight: 0.01                                                                          & 29.0           & 76.1          \\ \bottomrule
\end{tabular}
\label{tab:ablation}
\vspace{-4mm}
\end{table}

% \vspace{-2mm}

\subsection{Evaluation of converted acoustic features}
% (1) AC-classifier (2) Statistic comparison (3) F0-distribution
% Providing an objective evaluation of the converted acoustic feature before experimenting with ASR is meaningful. 
%It can reduce the number of trial and errors and give the guidance direction to design the conversion model. 
Since paired speech data are not available, i.e., parallel utterances of the same content are not available from adult and child domains, spectral distance measures like the Mel cepstral distortion (MCD)  are not applicable. An adult-child classification model is trained to distinguish the three domains, i.e., $A$, $C_1$ and $C_2$.  We hypothesize that the converted features are obtained from a high-quality $A2C_1$ conversion process if they are classified into the $C_1$ domain.
The percentages of different types of converted features classified as $C_1$ are shown as in Table \ref{tab:eval}. The \textit{CycleGAN} appears to perform very well, having $100\%$ converted features classified as $C_1$. However, the ASR model trained with these features shows performance degradation on test speech from $C_1$. Generally, DNN-based conversion methods are able to generate a high percentage of features classified as $C_1$. This may be related to the robustness issue of deep classification model. The pearson correlation coefficient between the WERs and $C_1$ classification percentages is $0.07$.  

\begin{figure}[t]
         \centering
         \includegraphics[width=\linewidth]{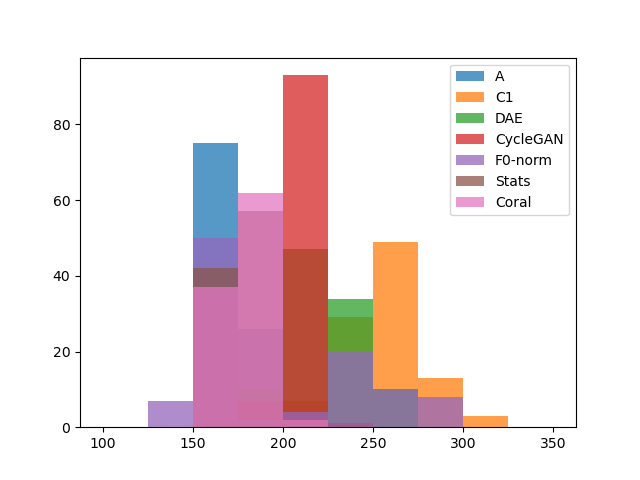}
         \caption{The F0 distribution of different acoustic features \\ (converted $A2C_1$ vs original $A$ \& $C_1$).}
         \label{fig:f0_dist}
\end{figure}

In view of distinctive F0 levels in adult and child speech, the utterance-wide median F0 values of these two domains are estimated.
%we estimate the median F0 values of each type of speech utterances.
The F0 distributions are visualized in Figure \ref{fig:f0_dist} using 100 utterances from different types of acoustic features, including the converted and the unconverted ones.  The ASR is expected to have better performance if the F0 distribution of the converted features is close to that of the $C_1$. The 1D Wasserstein distance is adopted to measure the discrepancy of two F0 distributions \cite{altschuler2017near}, which are listed in the last column of Table \ref{tab:eval}. The pearson correlation coefficient with WERs is $0.83$.

\begin{table}[t]
\centering
\caption{The objective evaluation of the converted features.}
\begin{tabular}{c|c|c|c}
\toprule
\begin{tabular}[c]{@{}c@{}}Conversion\\ Model\end{tabular} & \begin{tabular}[c]{@{}c@{}}WER (\%) \\ of $C_12C_1$\end{tabular} & \begin{tabular}[c]{@{}c@{}}Classified to \\ be $C_1$ (\%) \end{tabular} & \begin{tabular}[c]{@{}c@{}}F0 distance\\ to $C_1$\end{tabular} \\ \hline
None                                                       & 29.0                                                       & 0.0                                                        & 41.8                                                      \\ \hline
\textit{DAE}                                                         & 28.5                                                       & 95.7                                                       & 2.8                                                       \\
\textit{CycleGAN}                                                   & 31.7                                                       & 100.0                                                      & 126.9                                                      \\
\textit{F0-norm}                                                    & 28.3                                                       & 69.1                                                      & 5.5                                                       \\
\textit{Stats}                                                      & 30.3                                                       & 57.5                                                       & 19.7                                                       \\
\textit{Coral}                                                      & 30.3                                                       & 51.7                                                       & 19.7                                                       \\ \bottomrule
\end{tabular}
\label{tab:eval}
\vspace{-4mm}
\end{table}

\section{Conclusion}
In this paper, we compare the efficacies of different conversion methods to conduct adult-to-child conversion in the acoustic feature space. 
%The F0 normalization can achieve the best performance in terms of WER. 
The DAE-based conversion framework is investigated in detail with various settings, in which the DAT and F0-guided training are shown to be useful. In addition, using an adult-child deep classification model to judge the quality of conversion is less reliable. The distance of the converted feature set's F0 distribution to that of the target child domain presents a high correlation with the WER performance.
%the F0 distribution of the feature set is found to be an important attribute to reflect the conversion quality. The distance to the target child domain

%\newpage

\bibliographystyle{IEEEtran}

\bibliography{main}

\end{document}